\documentclass[fleqn,usenatbib]{mnras}

\usepackage{newtxtext,newtxmath}

\usepackage[T1]{fontenc}

\DeclareRobustCommand{\VAN}[3]{#2}
\let\VANthebibliography\thebibliography
\def\thebibliography{\DeclareRobustCommand{\VAN}[3]{##3}\VANthebibliography}

\usepackage{graphicx}	
\usepackage{amsmath}	
\usepackage[flushleft]{threeparttable}
\usepackage{multirow}
\usepackage{array}
\usepackage{xcolor}
\usepackage{ulem}

\newcommand{\ps}{\,s$^{-1}$}
\newcommand{\flux}{\,erg\,s$^{-1}$\,cm$^{-2}$}

\newcommand{\LxLb}{$L_\text{X}/L_\text{bol}$}
\newcommand{\nuLupi}{$\nu^2$\,Lupi}

\title[XUV-driven escape in TOI-431 \& \nuLupi]{The XUV-driven escape of the planets around TOI-431 \& \nuLupi}

\author[G. W. King et al.]{George W. King,$^{1,2,3}$\thanks{E-mail: kinggw@umich.edu}
L\'{i}a R. Corrales,$^1$
Jorge Fern\'{a}ndez Fern\'{a}ndez,$^{2,3}$
Peter J. Wheatley,$^{2,3}$
Isaac Malsky,$^1$
\newauthor Ares Osborn,$^{2,3}$
and David Armstrong$^{2,3}$
\\
$^{1}$Department of Astronomy, University of Michigan, Ann Arbor, MI 48109, USA\\
$^{2}$Department of Physics, University of Warwick, Gibbet Hill Road, Coventry, CV4 7AL, UK\\
$^{3}$Centre for Exoplanets and Habitability, University of Warwick, Gibbet Hill Road, Coventry, CV4 7AL, UK\\
}

\date{Accepted XXX. Received YYY; in original form ZZZ}

\pubyear{2024}

\begin{document}
\label{firstpage}
\pagerange{\pageref{firstpage}--\pageref{lastpage}}
\maketitle

\begin{abstract}
One of the leading mechanisms invoked to explain the existence of the radius valley is atmospheric mass loss driven by X-ray and extreme-ultraviolet irradiation, with this process stripping the primordial envelopes of young, small planets to produce the observed bimodal distribution. We present an investigation into the TOI-431 and \nuLupi\ planetary systems, both of which host planets either side of the radius valley, to determine if their architectures are consistent with evolution by the XUV mechanism. With \textit{XMM-Newton}, we measure the current X-ray flux of each star, and see evidence for a stellar flare in the TOI-431 observations. We then simulate the evolution of all of the transiting planets across the two systems in response to the high-energy irradiation over their lifetimes. We use the measured X-ray fluxes as an anchor point for the XUV time evolution in our simulations, and employ several different models of estimating mass loss rates. While the simulations for TOI-431\,b encountered a problem with the initial calculated radii, we estimate a likely short ($\sim$\,Myr) timespan for primordial envelope removal using reasonable assumptions for the initial planet. \nuLupi\,b is likely harder to strip, but is achieved in a moderate fraction of our simulations. None of our simulations stripped any of the lower density planets of their envelope, in line with prediction. We conclude that both systems are consistent with expectations for generation of the radius valley through XUV photoevaporation.
\end{abstract}




\section{Introduction}
The radius(-period) valley is a dearth of planets with radii of 1.5 to 2\,R$_\oplus$ at short orbital periods up to a few tens of days. First predicted almost a decade ago \citep{Owen2013,Lopez2013}, it was discovered observationally by the California-Kepler Survey following their reassessment of \textit{Kepler} system properties with improved stellar parameters \citep{Fulton2017}. Further analysis revealed a negative slope as a function of period \citep[e.g.][]{Martinez2019}, including that performed with a smaller sample of systems with more precise characterisation \citep{VanEylen2018}. The two populations of planets straddling the valley are thought to be the smallest planets able to retain a gaseous envelope (those just above the valley), and bare rock with at most a small atmosphere of heavier elements (those below the valley).

The two leading proposed mechanisms for producing the valley are core-powered mass loss and photoevaporative escape driven by X-ray/extreme-ultraviolet (EUV; together, XUV) light. Both of these produce the bare rock population via the stripping of primordial gaseous envelopes. In the core-powered mass loss mechanism, energy stored in the planet from its formation gradually radiates out into the envelope as the rocky core cools. The thermal energy supplied to the atmosphere in this way has been shown to be sufficient to drive substantial mass loss over the first Gyr or so of a planet's lifetime \citep{Ginzburg2016,Gupta2019,Gupta2020,Gupta2022}. These studies can reproduce both the bimodality and negative slope in radius-period space that are observed in the exoplanet population. Alternative explanations for the valley such as different core compositions for the two populations \citep{Venturini2020}, and a primordial generation at the epoch of gas accretion \citep{Lee2021,Lee2022} have also been proposed, and warrant further study. However, in this work, we focus on the XUV-driven escape mechanism.

\begin{table*}
\caption{Adopted stellar parameters for the two systems we investigated. Parameters are taken from \citet{Osborn2021} and \citet{Delrez2021} for TOI-431 and \nuLupi, respectively, except for distances which are from Gaia DR3 parallaxes \citep{GaiaDR3}.}
\label{tab:starParams}
\begin{tabular}{lcccccccc}
\hline \\[-0.25cm]
\multirow{2}{*}{Star} & \multirow{2}{*}{\begin{tabular}[c]{@{}c@{}}Spectral\\ Type\end{tabular}} & Distance         & $R_*$           & $M_*$         & $T_{\rm eff}$ & Age   & $P_{\rm rot}$ & $L_{\rm bol}$             \\
                      &                                                                          & (pc)             & (R$\odot$)      & (M$\odot$)    & (K)           & (Gyr) & (d)           & (L$\odot$)                \\[0.1cm]
\hline \\[-0.25cm]
TOI-431               & K5V                                                                      & $32.599\pm0.031$ & $0.729\pm0.022$ & $0.77\pm0.07$ & $4850\pm75$   & 1.9   & 30.5          & $0.264^{+0.023}_{-0.022}$ \\[0.05cm]
\nuLupi               & G3/5V                                                                    & $14.739\pm0.013$ & $1.058\pm0.019$ & $0.87\pm0.04$ & $5664\pm61$   & 12.3  & 23.8          & $1.038\pm0.059$          \\[0.1cm] 
\hline 
\end{tabular}
\end{table*}

\begin{table*}
\caption{Adopted planetary parameters for the transiting planets in the systems we investigated. Parameters are taken from \citet{Osborn2021} and \citet{Delrez2021} for the TOI-431 and \nuLupi\ systems, respectively. The numbers in brackets for the $P_{\rm orb}$ values are the uncertainties on the final two decimal places.}
\label{tab:planetParams}
\begin{tabular}{lcccccc}
\hline \\[-0.25cm]
\multirow{2}{*}{Planet} & $R_\oplus$                & $M_\oplus$              & $\rho$                 & $P_{\rm orb}$          & $a$                          & $T_{\rm eq}$ \\
                        & (R$_\oplus$)              & (M$_\oplus$)            & (g\,cm$^{-3}$)         & (d)                    & (au)                         & (K)          \\[0.1cm]
                        \hline \\[-0.25cm]
TOI-431\,b              & $1.28\pm0.04$             & $3.07^{+0.35}_{-0.34}$  & $8.0\pm1.0$            & 0.490047($^{10}_{07}$) & $0.0113^{+0.0002}_{-0.0003}$ & $1862\pm42$  \\[0.05cm]
TOI-431\,d              & $3.29^{+0.09}_{-0.08}$    & $9.9\pm1.5$             & $1.36\pm0.25$          & 12.46103(02)           & $0.098\pm0.002$              & $633\pm14$   \\[0.05cm]
\nuLupi\,b              & $1.664\pm0.043$           & $4.72\pm0.42$           & $5.62^{+0.72}_{-0.66}$ & 11.57797($^{08}_{13}$) & $0.0964\pm0.0028$            & $905\pm14$   \\[0.05cm]
\nuLupi\,c              & $2.916^{+0.075}_{-0.073}$ & $11.24^{+0.65}_{-0.63}$ & $2.49^{+0.25}_{-0.23}$ & 27.59221(11)           & $0.1721\pm0.0050$            & $677\pm11$   \\[0.05cm]
\nuLupi\,d              & $2.562^{+0.088}_{-0.079}$ & $8.82^{+0.93}_{-0.92}$  & $2.88^{+0.43}_{-0.40}$ & 107.245(50)            & $0.425\pm0.012$              & $431\pm7$   \\[0.1cm]
\hline
\end{tabular}
\end{table*}

In the XUV-driven escape mechanism, XUV photons are absorbed high in the atmosphere of the planet, efficiently heating this region, which sets up and drives a hydrodynamic outflow. The XUV irradiation is highest in the early life of the system and reduces in intensity as the star undergoes magnetic braking \citep[e.g.][]{Micela1985,Guedel1997,Micela2002,Feigelson2004,Jackson2012,Johnstone2021}, so one expects the resulting mass loss to also reduce as the system ages. The first 100\,Myr, when the XUV irradiation is at its highest level, has therefore been widely thought to be the dominant epoch for evolution due to XUV photons, and the timescale on which the radius valley is imprinted into the population \citep[e.g.][]{Lopez2013,Lammer2014,Owen2017}. This would provide a possible method of distinguishing which mechanism, if either, dominates the generation of the radius valley, if one had a large enough population of exoplanets at all young ages up to 1\,Gyr. However, \citet{EUVevolution} showed that the reduction in EUV irradiation may be significantly slower than X-rays. This in turn could prolong the timescale on which sub Neptune-sized planets are significantly evolved via XUV-driven escape up towards that expected for core-powered mass loss. Other methods of trying to distinguish between the mechanisms have also been proposed, such as examining the radius valley in a 3D parameter space \citep[][]{Rogers2021}.

The NASA \textit{TESS} mission has discovered many multi-planet systems over the past few years which contain at least one planet just above the radius valley, and at least one planet just below it \citep[e.g.][]{Guenther2019,Cloutier2020,Demory2020,Hawthorn2022}. Such systems are ideal laboratories to test the mechanisms theorised to explain the valley by examining whether they have properties consistent with a particular mechanism(s). In the case of photoevaporation, one advantage of having the planets in the same system is that they will have experienced an identically evolving XUV irradiation history \citep[e.g.][]{Owen2020}. 

Among the \textit{TESS} discoveries of systems with planets either side of the radius valley are TOI-431 and \nuLupi\ (HD\,136352), both of which host three known planets. Two of the TOI-431 planets have been observed to transit: b and d \citep{Osborn2021}. The system is unusual in that planet d transits despite being exterior to the non-transiting planet c. Planet b orbits 
on a 0.5\,d period, and its density is consistent with it being a rocky core with no envelope. Planet c has a minimum mass similar to b, while planet d has a density consistent with having a gaseous envelope. The three planets known to orbit the naked-eye brightness \nuLupi\ star were originally discovered via radial velocities \citep{Mayor2011,Udry2019}. Transits of planets b and c were later detected by \textit{TESS} \citep{Kane2020}, and d was also found to transit by \textit{CHEOPS} \citep{Delrez2021}. Planet b is consistent with being a rocky core, while planets c and d have densities that suggest they have a gaseous envelope. The visual brightness of \nuLupi\ means that the system is an excellent target for atmospheric characterisation of its planets with transmission spectroscopy. Furthermore, with both systems residing relatively nearby to the Solar System, they are possible future targets for Ly-$\alpha$ or helium triplet observations targeting escaping atmospheres. In our work, we adopt parameters for the TOI-431 system from \citet{Osborn2021}, and for the \nuLupi\ system from \citet{Delrez2021}. We display some key characteristics for the stars and planets in Tables \ref{tab:starParams} and \ref{tab:planetParams}, respectively.

The aim of this work is to test whether the TOI-431 and \nuLupi\ systems are consistent with the XUV-driven photoevaporation mechanism for generating the radius gap. We measured the X-ray emission from the two stars with \textit{XMM-Newton}, which is important for two reasons. First, knowledge of the XUV irradiation is an important input to interpreting observations attempting to detect extended or escaping atmospheres of the planets in these systems. Second, direct measurement is important for systems of interest as empirical relations that can be used to estimate X-ray emission are subject to variation 
up to an order of magnitude \citep[e.g.][]{Wright2011,Wright2018,Jackson2012}. Following analysis of our observations, we ran simulations for the lifetime evolution of the transiting planets in the systems under the action of XUV-driven photoevaporation, using the measured fluxes as an anchor point for the escape-driving XUV emission evolution. Using the simulations, we assessed whether XUV irradiation could have sculpted the systems into what we observe today.

\section{Observations}
\label{sec:obs}

We observed TOI-431 and \nuLupi\ with \textit{XMM-Newton} on 2021 Aug 28 and 2021 Sep 12, respectively (OBSIDs 0884680101 and 0884680201; PI: King). In both cases, the prime instrument was the EPIC-pn camera.

We operated the simultaneous Optical Monitor (OM) observations in fast mode with a single near-ultraviolet (NUV) filter, in order to monitor the stars for short term variation, such as flares. The TOI-431 observations were made with the UVW1 filter (effective wavelength 291\,nm, width 83\,nm), while the \nuLupi\ observations were made with UVW2 (effective wavelength 212\,nm, width 50\,nm).

Both targets were successfully detected in the EPIC-pn observations. We analysed the observations using the Scientific Analysis System (\textsc{sas} 19.1.0), using the standard process, as detailed on the ``SAS Threads" webpage\footnote{\url{http://www.cosmos.esa.int/web/xmm-newton/sas-threads}}. Elevated background due to Solar protons, a problem well-known to affect \textit{XMM-Newton} observations \citep{Walsh2014}, affected small portions of both observations in the EPIC-pn camera. As we have done previously \citep[e.g.][]{SurveyPaper}, we double the threshold suggested in the SAS Threads for cutting events due to high background, as this results in less cuts, and therefore more usable counts, without adversely affecting the results. For TOI-431, there were two high background peaks in the final quarter of the observation, while the \nuLupi\ observations contain a single peak right at the beginning of the observation. Both EPIC-MOS cameras were unaffected by this issue throughout both observations. 

In our data reduction of the OM data, we used the same processes as described in \citet{SurveyPaper}: following running the standard chains for image and fast mode, we corrected the fast mode time series using the image mode count rates. Both observations exhibit relatively high count rates of over 100\,\ps, though both are well within the valid range for coincidence loss correction, given the frame times of 7.5\,ms and 5.5\,ms for TOI-431 and \nuLupi, respectively. A description of the brightness limits of the OM can be found in section 3.5.5 of the \textit{XMM-Newton} Users Handbook\footnote{Available at \url{https://heasarc.gsfc.nasa.gov/docs/xmm/uhb/omlimits.html}}; the raw fast mode count rates (i.e. before corrections for e.g. coincidence loss) for comparison with table 21 and the surrounding descriptions are 60\,\ps and 70\,\ps for TOI-431 and \nuLupi, respectively.

Both sets of OM images exhibit commonly seen artifacts. For TOI-431, there was a faint, extended loop of emission extending from the bottom right of the location of the target. This is caused by a chamfer in the housing of the detector reflecting light from a bright source just outside the detector field of view \citep{Mason2001}. The overall contamination of the target from this effect is low. The number of counts per pixel in the loop is much less than one per cent that in the central region of the target's point spread function (PSF), and the majority of the loop does not overlap the source. For \nuLupi, there is a faint signature of the detector's modulo-8 pattern in the source PSF \citep{Mason2001}. The algorithm that removes this signature from the background of OM images is imperfect for bright sources due to coincidence losses, though photometery is unaffected due to the number of photons being conserved \citep{Breeveld2010}.

\section{\textit{XMM-Newton} results}

\subsection{X-ray and NUV light curves}

\begin{figure}
\centering
 \includegraphics[width=\columnwidth]{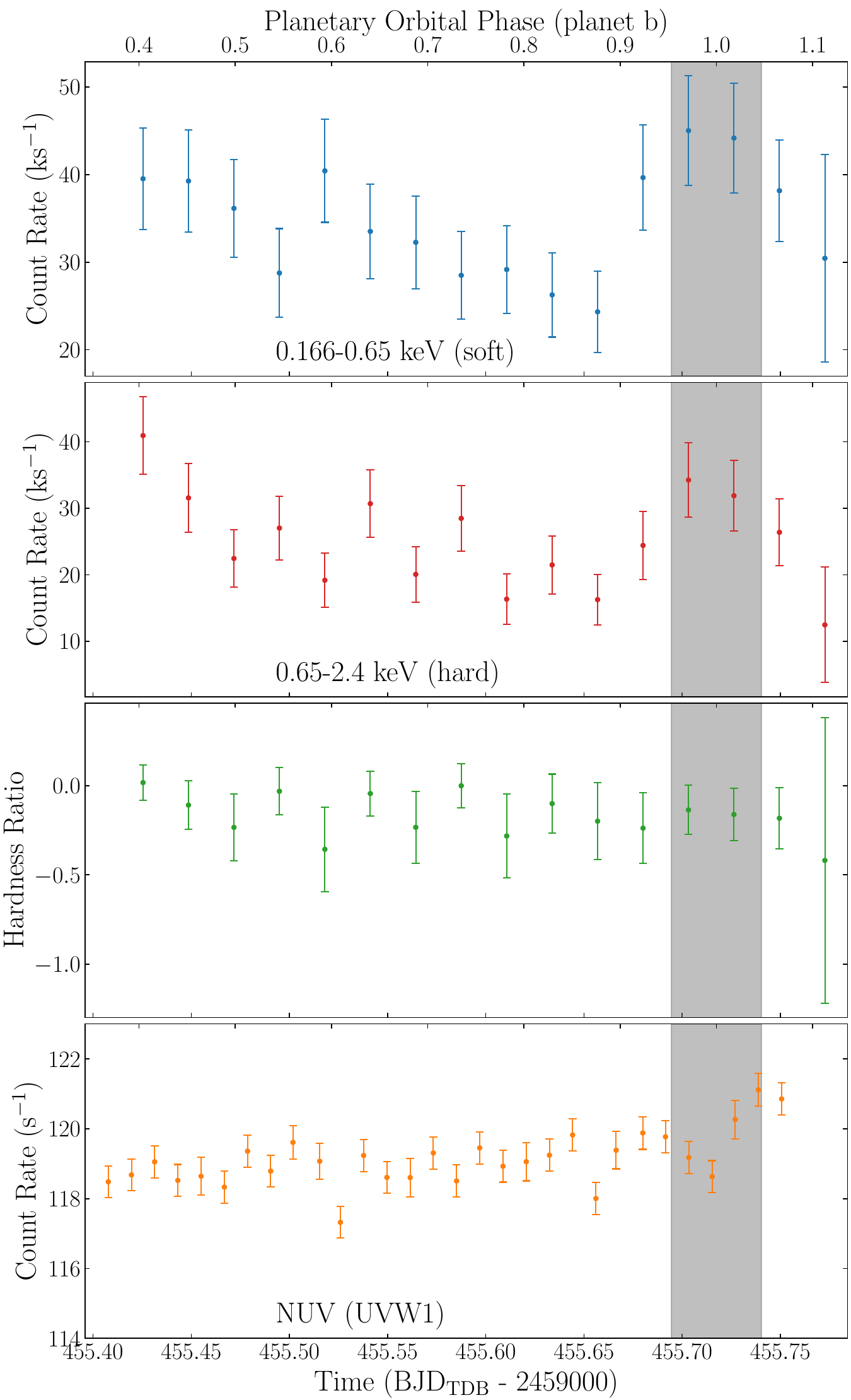}
 \caption{X-ray and NUV light curves for TOI-431. The top and upper middle panels show the soft (0.166--0.65\,keV) and hard (0.65--2.4\,keV) energies, respectively. The lower middle panel is the hardness ratio (see main text for definition). All three panels are based on the count rate data from all three EPIC cameras together, binned to a cadence of 2\,ks. The bottom panel shows the NUV light curve measured with the OM, using the UVW1 filter. The shaded grey region shows the phases covered by the optical transit of planet b.}
 \label{fig:t431_xlc}
\end{figure}

\begin{figure}
\centering
 \includegraphics[width=\columnwidth]{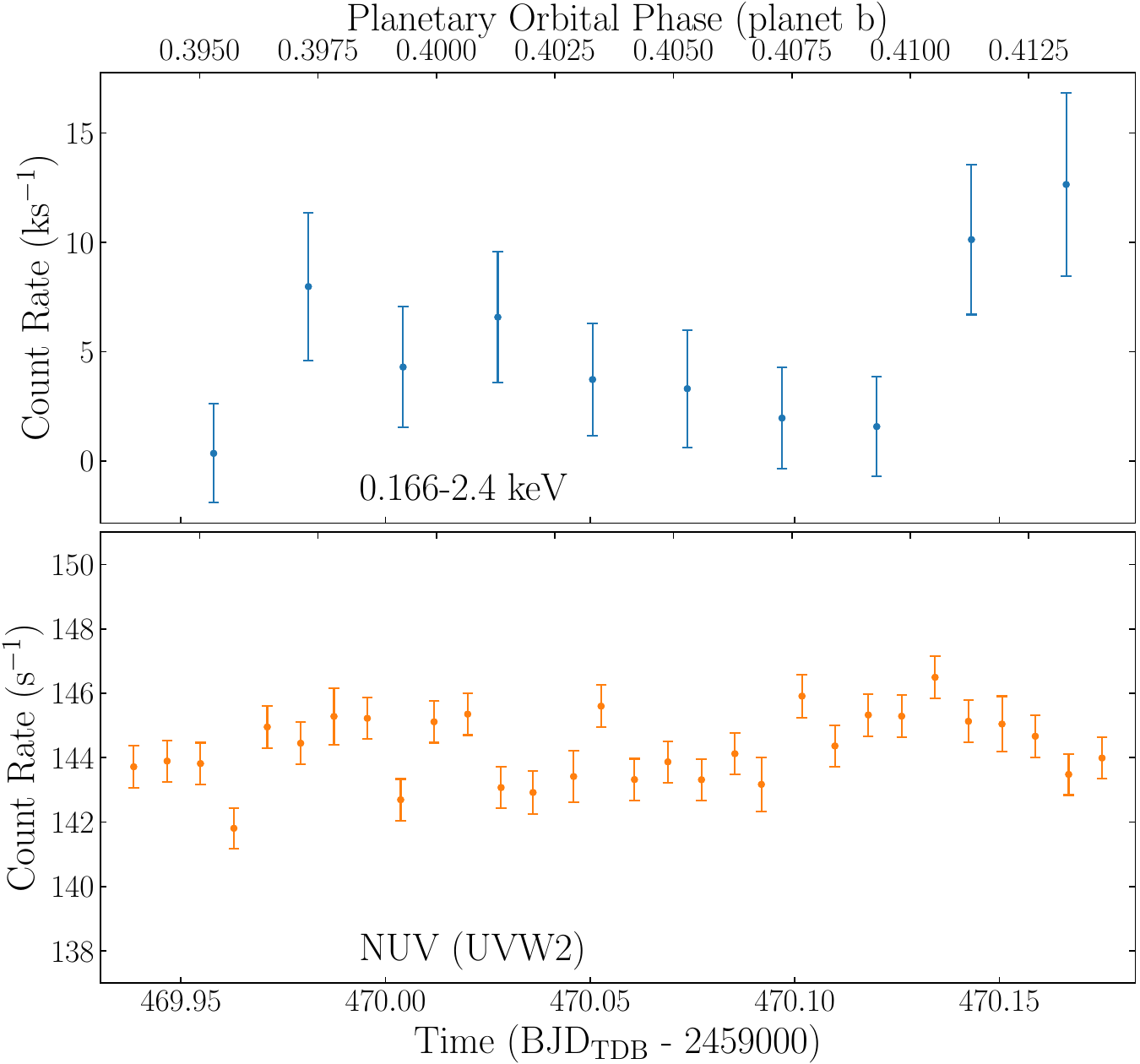}
 \caption{X-ray and NUV light curves for \nuLupi. The X-ray data are plotted in the top panel, binned to a cadence of 2\,ks, and for the full 0.166--2.4\,keV energy band. They are the total from coadding all three EPIC cameras together. In the bottom panel is the NUV data measured with the OM, using the UVW2 filter.}
 \label{fig:nuLupi_xlc}
\end{figure}

In Figs.~\ref{fig:t431_xlc} we show the observed \textit{XMM-Newton} light curves for TOI-431, and for \nuLupi\ in Fig.~\ref{fig:nuLupi_xlc}. The X-ray light curves depict the coadded count rates across the three EPIC cameras - pn, MOS1, and MOS2 - though for both targets the observed counts are dominated by EPIC-pn. Also included are the NUV light curves plotted from the data taken with the OM.

The TOI-431 X-ray light curve in Fig. \ref{fig:t431_xlc} is split across the top two panels, separating the soft (0.166--0.65\,keV) and hard (0.65--2.4\,keV) energies. The orbital phases plotted are for planet b, where mid-transit is at a phase of 1. Though the observation covers a primary transit (grey shaded region), there is no transit visible in the X-ray light curve. This is unsurprising given the small size of planet b and the count rate errors at this cadence of 10--20 per cent. The phases covered by the observation for the other transiting planet d are 0.437--0.466.

The light curves in both energy ranges show evidence of stellar X-ray variation. During the last quarter of the observation, the count rate increases by about 50 per cent on the previous quarter in both the soft and hard bands. While there are two peaks in the high-energy background due to Solar protons (discussed in Section~\ref{sec:obs}) around 24 and 30\,ks into the observation, these are short, lasting less than a ks in each case. Additionally, both high background peaks occur in the ramp up and down, and thus cannot explain the peak region of elevated count rate. The increase is similar in both energy bands. We test this further by plotting the hardness ratio, $HR$, light curve in the lower middle panel of Fig \ref{fig:t431_xlc}. We use the following definition of hardness ratio:
\begin{equation}
    HR = \frac{H - S}{H + S},
    \label{eq:HR}
\end{equation}
where $S$ and $H$ are the soft and hard band count rates, respectively. The resulting plot shows no evidence of variation in the hardness ratio throughout the observation, including during the possible flaring epoch. The fact that the increase is similar in both energy bands point to the variation possibly not being the result of a flare on the host star, since stellar flares typically shows a hardening of the X-ray emission \citep[e.g.][]{Reale2001,Telleschi2005,Pye2015}. However, the NUV light curve in the bottom panel of Fig. \ref{fig:t431_xlc} also shows an increase in count rate at the end of the observation. In this case, the rise is 2\,\ps (1.7 per cent), and appears to peak some 50 minutes after the start of the bump in X-rays, though there is some overlap between the increases. The two highest points in the X-ray light curve cover the phase range 0.947 -- 1.042, while including the point either of this extends the range to 0.900 to 1.089. In comparison, the phase range covered by the last three points in the OM light curve constituting the rise is 1.008 -- 1.080. Additionally, delays in the peak of flares between different wavelengths on the order of tens of minutes would not be unprecedented \citep[e.g.][]{MacGregor2021}. It is unfortunate that the OM observation ended when it did, because observations of the following hour could have allowed us to unambiguously identify whether or not it was a true stellar flare by looking for the characteristic flare decay shape. In its absence, and in the absence of a hardening of the X-ray emission, we can only conclude the ramp up to possibly be due to a stellar flare, with other scenarios possible too. One possibility is the rotation into view of a particularly active region of star. In that instance, differing optical thickness between X-ray and NUV could explain the delay in the increase between the two wavelengths.


We do not split the \nuLupi\ X-ray light curve in the top panel of Fig.~\ref{fig:nuLupi_xlc} into separate soft and hand bands, as there were too few hard energy photons through the observation for it to be useful on its own. The observation did not cover a transit of any of the three planets. Planet b's orbital phases during the observation are plotted on Fig.~\ref{fig:nuLupi_xlc}. For planets c and d, the orbital phases were 0.6841--0.6925 and 0.2908--0.2930, respectively.


As for TOI-431, the \nuLupi\ light curve also shows evidence of stellar variation. The most notable of this is in the last 3\,ks of the observation, where the count rate jumps up from being consistent with zero to over 12.5\,ks$^{-1}$. This could be the beginning of a stellar flare, but again, 
without capturing the full flare in the observation, it is difficult to speculate further. With an age of 12.3\,Gyr, one might not expect to observe many flares from this star, as flare occurrence decreases with age \citep[e.g][]{Skumanich1986,Davenport2019}. Additionally, the NUV light curve with the OM shows no rise at this point in the observation, and instead appears to decrease slightly. Overall, the OM light curve is largely unremarkable and exhibits little variability.

\subsection{X-ray Spectra}
\label{ssec:Xspec}

\begin{figure}
\centering
 \includegraphics[width=0.88\columnwidth]{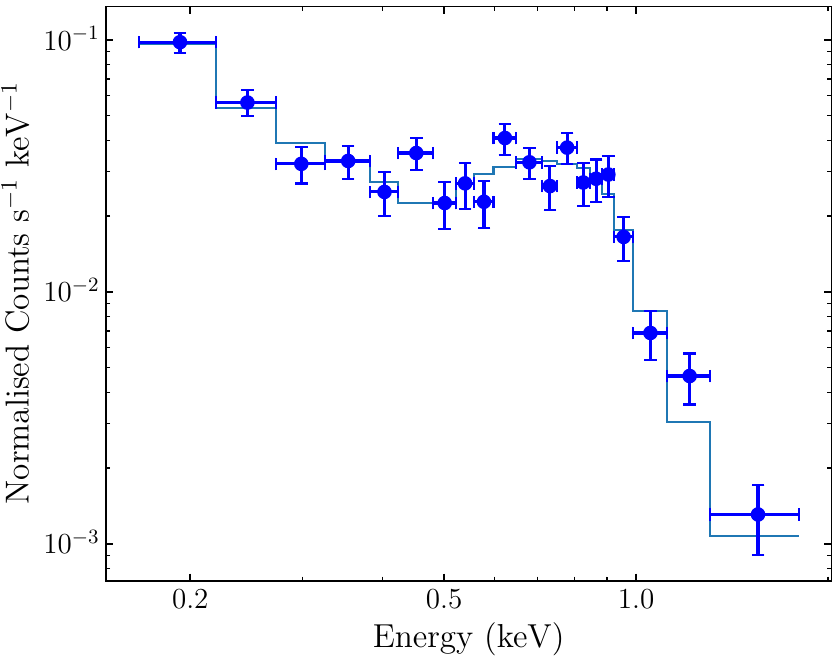}
 \caption{X-ray spectrum for TOI-431, observed by the EPIC-pn camera. For details of the overlaid model fit, see the main text.}
 \label{fig:t431_spec}
\end{figure}

\begin{figure}
\centering
 \includegraphics[width=0.88\columnwidth]{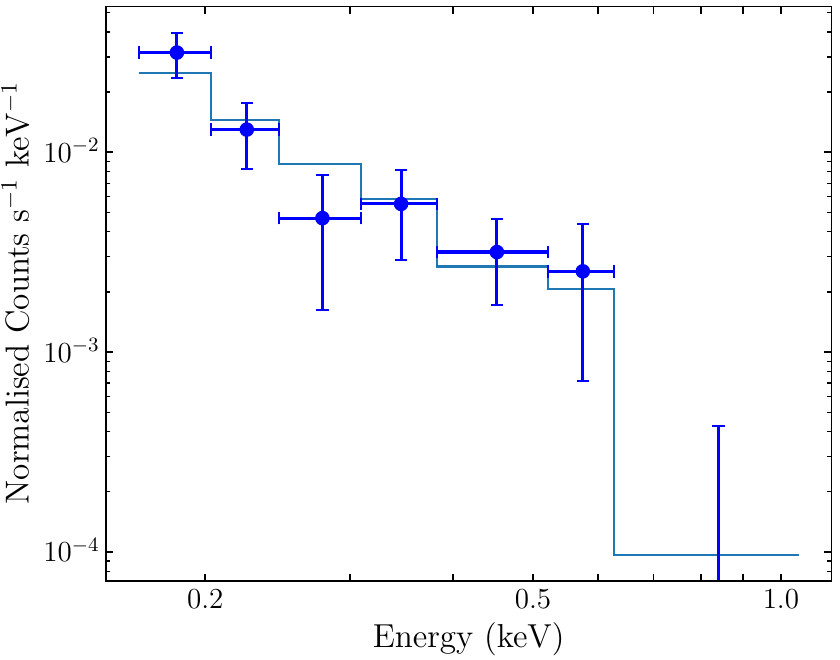}
 \caption{As Fig.~\ref{fig:nuLupi_spec} but for \nuLupi. The last point on the right for which only the top of errorbar is visible has a value $(-0.37\pm4.63)\times10^{-4}$, and so cannot be shown on the log axis. }
 \label{fig:nuLupi_spec}
\end{figure}

\begin{table}
\centering
\caption{Best fit temperatures, emission measures, and unabsorbed fluxes at Earth for the fits to our X-ray spectra of TOI-431 and \nuLupi. We also give luminosities and fluxes at the planets calculated using from $F_{\rm x}$.}
\label{tab:fluxes}
\begin{threeparttable}
\begin{tabular}{lccl}
\hline
Parameter      & TOI-431                                           & \nuLupi                                            & Units         \\ \hline
\\[-0.25cm]
$kT_1$          & $0.080^{+0.015}_{-0.031}$                         & $0.0847^{+0.0056}_{-0.0392}$                      & keV           \\[0.05cm]
EM$_1$          & $\left(4.4^{+15.3}_{-1.2}\right)\times10^{50}$    & $\left(4.11^{+42.4}_{-0.76}\right)\times10^{50}$  & cm$^{-3}$     \\[0.05cm]
$kT_2$          & $0.255^{+0.024}_{-0.031}$                         & -                                                 & keV           \\[0.05cm]
EM$_2$          & $\left(1.87^{+0.45}_{-0.26}\right)\times10^{50}$  & -                                                 & cm$^{-3}$     \\[0.05cm]
$kT_3$          & $0.775^{+0.102}_{-0.071}$                         & -                                                 & keV           \\[0.05cm]
EM$_3$          & $(7.8\pm2.0)\times10^{49}$                        & -                                                 & cm$^{-3}$     \\[0.05cm]
$F_{\rm X}$     & $\left(6.35^{+0.17}_{-0.45}\right)\times10^{-14}$ & $\left(1.20^{+0.25}_{-0.28}\right)\times10^{-14}$ & \flux          \\[0.05cm]
$L_{\rm X}$     & $\left(8.07^{+0.21}_{-0.58}\right)\times10^{27}$  & $\left(3.12^{+0.64}_{-0.74}\right)\times10^{26}$  & erg\,s$^{-1}$ \\[0.05cm]
$L_{\rm EUV}$   & $\left(2.47^{+0.26}_{-0.35}\right)\times10^{28}$  & $\left(6.8^{+1.5}_{-1.7}\right)\times10^{27}$     & erg\,s$^{-1}$ \\[0.05cm]
$F_{\rm XUV,b}$ & $91200^{+7400}_{-9900}$                           & $274^{+57}_{-65}$                                 & \flux          \\[0.05cm]
$F_{\rm XUV,c}$ & $4300^{+350}_{-460}$                              & $86^{+18}_{-20}$                                  & \flux          \\[0.05cm]
$F_{\rm XUV,d}$ & $1210^{+100}_{-130}$                              & $14.1^{+3.0}_{-3.4}$                              & \flux          \\[0.05cm] \hline
\end{tabular}
\begin{tablenotes}
\item Note that we use the following energy band definitions:
\item X-ray: 0.2--2.4\,keV
\item EUV: 0.0136--0.2\,keV
\end{tablenotes}
\end{threeparttable}
\end{table}

We fit the EPIC-pn observations in \textsc{Xspec} 12.11.1. We did not include the EPIC-MOS data in the fits for either star because the signal was small compared to the EPIC-pn data, and so its inclusion did not improve our model fits.

The TOI-431 data was binned to a minimum of 25 counts per bin, while the \nuLupi\ data was binned to 10 counts per bin due to the lower number of overall counts. We fit the spectra with APEC models, which describe an optically thin, collisionally ionised plasma \citep{Smith2001}. For TOI-431, we had to use three temperature components to obtain a good fit, whereas for \nuLupi\ a single temperature was sufficient due to the much fewer spectral counts. We also included a TBABS term to account for interstellar absorption \citep{Wilms2000}, fixing the hydrogen column density to an assumed local density of 0.1 cm$^{-3}$ \citep{Redfield2000}. Abundances were fixed at Solar values, according to \citet{Asplund2009}. We used Cash statistics in our fitting \citep{Cash1979}, and used \textsc{Xspec}'s built-in MCMC sampler in order to assess the 1-$\sigma$ uncertainties on the model parameters and fluxes.

Figures~\ref{fig:t431_spec} and \ref{fig:nuLupi_spec} show our extracted spectra and best fitting models for TOI-431 and \nuLupi, respectively, while in Table \ref{tab:fluxes} we give our best fit model temperatures, $kT$, emission measures, EM, and unabsorbed fluxes at Earth, $F_{\rm X}$. Using our fluxes, we compute the X-ray luminosity, $L_{\rm X}$, of the stars, and combine this with the X-ray-EUV scaling relations from \citet{SurveyPaper} to estimate the EUV luminosities, $L_{\rm EUV}$. Finally, we estimate the current XUV irradiation rate at each of the planets in the two systems by scaling to their semi-major axes.

Both spectra are dominated by softer energies, especially in the case of the old star \nuLupi\ for which the detected emission above 0.6\,keV is consistent with zero. This is not so surprising since the coronae of older stars have been seen to exhibit much lower emission above this energy compared to younger stars \citep[e.g.][]{Guedel1997}, and given the observed relationship between the emission measure weighted average coronal temperature and X-ray surface flux (which in turn decreases with age) \citep{JohnstoneGuedel2015}. For TOI-431 however, there is a small bump in the spectrum between 0.5 and 0.9\,keV, characteristic of Fe L-shell emission. Unlike for much younger stars where this bump can dominate, 
the overall spectrum is still dominated by the soft end, suggestive of a star that is $\gtrsim 1$\,Gyr in age. This is consistent with the previously inferred age of $1.9\pm0.3$\,Gyr \citep{Osborn2021}, estimated from its chromospheric activity data and the \citet{Mamajek2008} empirical relations.





\section{Atmospheric Evolution Simulations}
\label{sec:sims}
Using our measured fluxes as an anchor point, we ran a series of simulations to investigate the evolutionary histories of the five transiting planets in the two systems in our sample. We do not consider TOI-431\,c as it is not transiting, and as such we have no measurement of the current radius of the planet. 


For the planetary evolution, we employed the 1-D stellar evolution code Modules for Experiments in Stellar Astrophysics \citep[MESA,][]{Paxton2011,Paxton2013,Paxton2015,Paxton2018,Paxton2019,Jermyn2023} version 12778. The code we implemented is exactly that used by \citet{Malsky2020} and \citet{Malsky2023}, based on \citet{Chen2016}, except for the final step in which the planet is evolved in time with XUV irradiation. We edited this final step to change the mass loss rate prescription, for which we use three different methods described below, and the time evolution of the XUV emission from the star, which we calculated using the Mors code \citep{Johnstone2021}.

The evolutionary tracks calculated by the \textsc{Mors} code provide a more complex XUV luminosity progression than assuming a constant saturated period followed by a power law decline \citep[e.g.][]{Jackson2012}. For stars a few hundred Myr old, there is evidence of increasing dispersion in the observed \LxLb\ values, which later reconverge at an age of about a Gyr \citep[see also e.g.][]{Tu2015}, something which \textsc{Mors} aims to model. We generated an evolution track for each star based on the stellar mass, then scaled the median track to the measured $L_{\rm XUV}$ from Section \ref{ssec:Xspec}. We note that the measured $L_{\rm XUV}$ for \nuLupi\ was a factor of three smaller than the median track calculated by \textsc{Mors}. This could be related to the relatively uncertain age of the star, it being in a quiet activity state during our observations, or just a demonstration of the scatter in the XUV time evolution relations. We also scaled the \textsc{Mors} $L_{\rm bol}$ track for both stars to their measured values, although the effect on the simulations from this correction is very small. 

We adopt the same atmospheric boundary conditions as \citet{Malsky2023}: a grey Eddington T($\tau$) relation, and model the atmosphere up to an optical depth of 2/3. Following \citet{Malsky2020}, we calculate the incident bolometric flux and the column depth of irradiation for each planet. Furthermore, we use the gaseous mean opacity tables from \cite{Freedman2014}.

\begin{table}
\centering
\caption{Choices for the starting mass and envelope mass fraction grid for the planets in the simulations.}
\label{tab:initGrid}
\begin{threeparttable}
\begin{tabular}{lcccccc}
\hline
Planet     & \multicolumn{3}{c}{Mass (M$_\oplus$)} & \multicolumn{3}{c}{Envelope Mass Fraction} \\
           & Lowest      & Highest    & Step       & Lowest        & Highest     & Step         \\ \hline
TOI-431\,b$^*$ & 2.25         & 7          & 0.25        & 0.01         & 0.40        & 0.01        \\
TOI-431\,d & 9           & 14         & 0.25        & 0.05          & 0.25        & 0.02      \\
\nuLupi\,b$^*$ & 4.0         & 9.5        & 0.5        & 0.01         & 0.39        & 0.01        \\
\nuLupi\,c & 11          & 17.5       & 0.25        & 0.05          & 0.30        & 0.02         \\
\nuLupi\,d & 8           & 14         & 0.25        & 0.05          & 0.25        & 0.02         \\ \hline
\end{tabular}
\begin{tablenotes}
\item $^*$ Note that for TOI-431\,b and \nuLupi\,b, we only use combinations of the starting mass and envelope mass fraction grids listed that imply a core mass within 3-$\sigma$ of the current measured mass of the planet.
\end{tablenotes}
\end{threeparttable}
\end{table}

In \citet{PraesepePaper}, we predicted the mass loss rates and future evolutionary tracks for four transiting planets in the 670~Myr old Praesepe open cluster. This time, we focus on reconstructing the past mass loss history for older field age systems. This meant defining a grid of starting masses and envelope mass fractions, displayed for each planet in Table \ref{tab:initGrid}. The grid choices were based on the current known mass characteristics of the planets. The range of simulated masses are wider for the two planets consistent with being bare rock at the present age, TOI-431\,b and \nuLupi\,b, since the aim of those simulations are somewhat different, as explained in Section \ref{ssec:stripped}.

We employed a few different methods of estimating the mass loss rates as part of our simulations, in order to compare the methods. The first of these is energy-limited mass loss, first formulated in the context of the Solar System planets by \citet{Watson1981}, and which has since seen widespread use for exoplanets over the past two decades \citep[e.g.][]{LDE2007,SanzForcada2011,Salz2015,Louden2017,Poppenhaeger2021}. The energy-limited mass loss rate, $\dot{M}_{\rm En}$, is given by \citep{Baraffe2004,Erkaev2007}
\begin{equation}
\dot{M}_{\rm En} = \frac{ \beta^2 \eta \pi F_{\rm XUV} R^3_{\rm p} }{ G K M_{\rm p} },
\label{eq:enML}
\end{equation}
where $\eta$ is the heating efficiency, $\beta$ is a correction factor to the distance above the optically-measured radius of the planet at which XUV photons are on average absorbed, and $K$ is a Roche-lobe correction factor \citep{Erkaev2007}. We assume a canonical $\eta$ value of 0.15 \citep[e.g.][]{Watson1981,Salz2015,Kubyshkina2018}, and a $\beta$ value of 1. The latter is in reality a lower limit, as XUV photons will be typically absorbed higher up in the planet's atmosphere.

The second method of estimating mass loss we explored was the \textsc{Ates} photoionisation hydrodynamics code \citep{Caldiroli2021}. 
Performing the full hydrodynamic calculation for each step on each track would not have been computationally possible. We do, however, run \textsc{Ates} in order to obtain current mass loss rates for the planets in Section \ref{ssec:currentML}. To simulate the grid of mass loss histories within a reasonable timeframe, we instead use an approximation to their code provided in a follow up paper \citep{Caldiroli2022}. The approximation gives both the mass loss rate and effective efficiency function, $\eta_{\rm eff}$, based on fitting an analytic expression to the results of the \textsc{Ates} code. Note that $\eta_{\rm eff}$ is not the same at the heating efficiency $\eta$ defined above, as $\eta_{\rm eff}$ also encapsulates the effects of the unknown $\beta$ parameter.

The final method was based upon the hydrodynamic modelling of \citet{Kubyshkina2018}. 
They produced a grid of models to estimate the mass loss rate, given some star and planet characteristics. As with ATES, the interpolator they provide in \citet{Kubyshkina2021} was too computationally expensive to couple with MESA, and so we instead use their ``hydro-based approximation." This was detailed was described in \citet{Kubyshkina2018HBA}, with the relation we used given in their equation 9. This is an analytic expression fitted to values they obtained from their grid of models.

We began our evolution models at 5\,Myr old, an age by which the majority of protoplanetary discs have dissipated \citep[e.g.][]{Haisch2001,Mamajek2009,Ribas2015}. We do not directly consider the ``boil-off" phase in our models, which is a more rapid process thought to occur on kyr to Myr time-scales immediately after the disc dissipates, and before the onset of meaningful XUV photoevaporation \citep[e.g.][]{Stokl2015,OwenWu2016,Ginzburg2016,Fossati2017,Rogers2024}. During this phase, the accumulated planetary atmosphere can contract significantly. The release of binding energy during contraction, together with the atmosphere no longer being protected from the bolometric stellar radiation by the disc, can drive substantial mass loss in this short epoch. We note that the \citet{Kubyshkina2018} model does partially cover this effect, and so our results based on that work could be affected by this phase of evolution.



\subsection{Results for Gaseous Planets}
\label{ssec:gaseous}




\begin{figure}
\centering
 \includegraphics[width=\columnwidth]{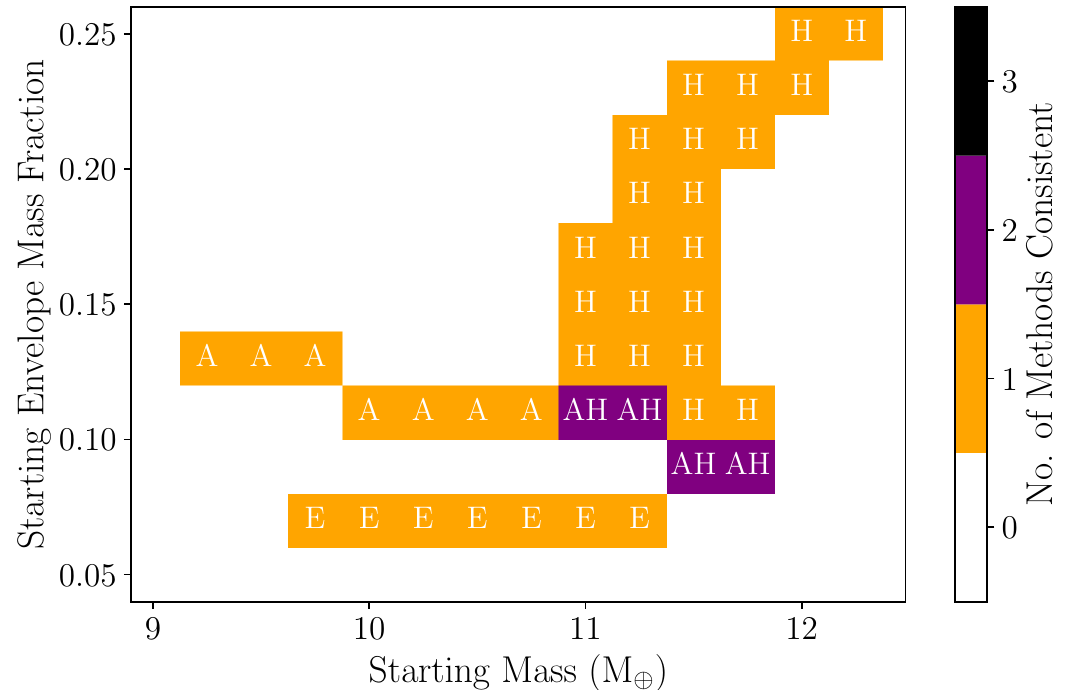}
 \caption{Heat map for the TOI-431\,d simulations, displaying the starting conditions which result in evolutionary tracks where the radius and mass are consistent with the measured values to within 1-$\sigma$ at the current estimated age. The squares are coloured according to the number of mass loss calculation methods that produce a consistent result. The labels display which method(s) produced a consistent result: E - energy limited; A - \textsc{Ates} approximation; 
 H - analytic hydro-based approximation to the \citet{Kubyshkina2018} models \citep{Kubyshkina2018HBA}. Some simulation tracks lie outside the limits of the plot, but none of those produced any consistent tracks.}
 \label{fig:heatmap_toi_d}
\end{figure}

\begin{figure*}
\centering
 \includegraphics[width=\textwidth]{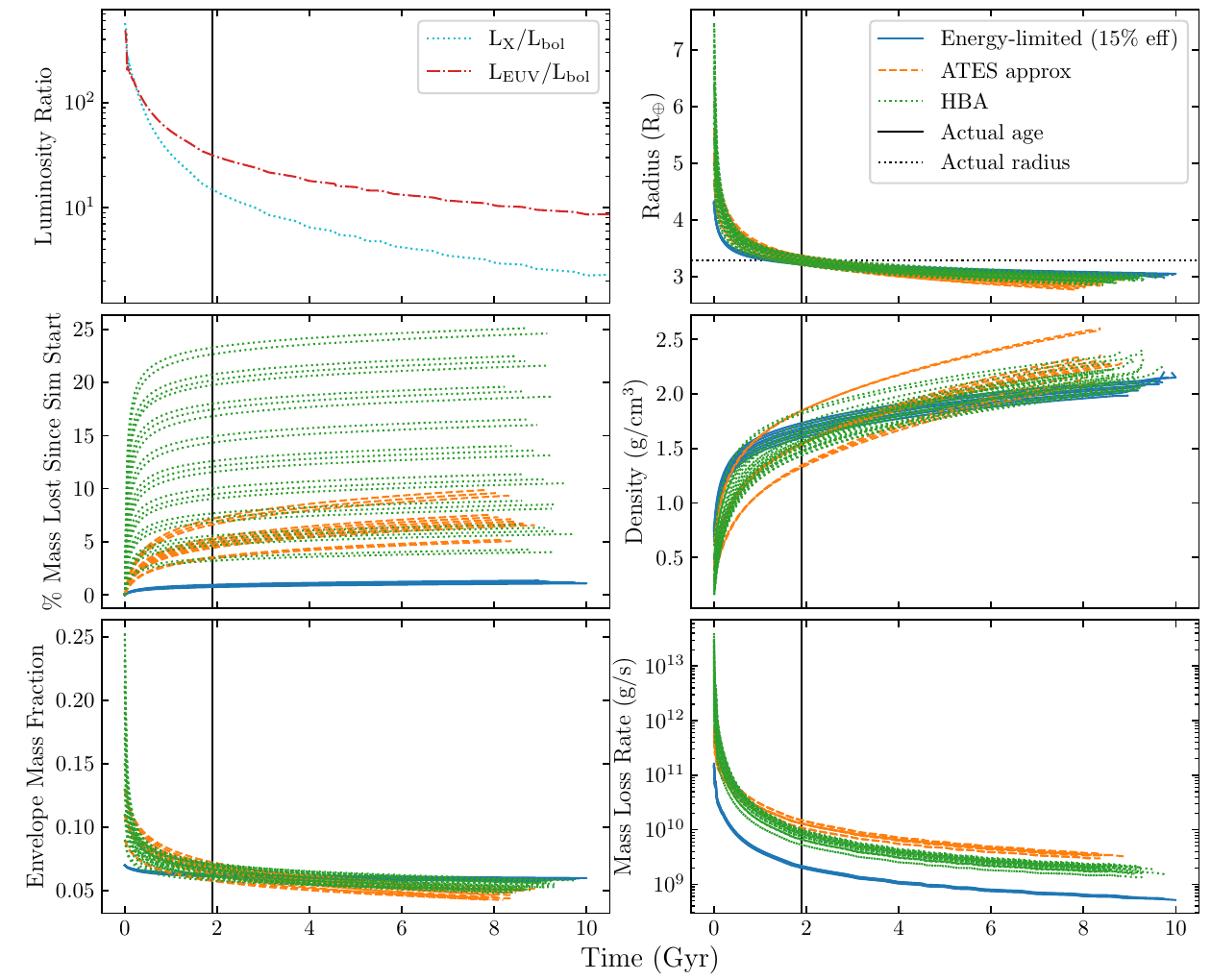}
 \caption{The subset of evolutionary tracks for TOI-431\,d where the planet's radius (marked as a black dotted line in the top right panel) and mass matched the measured values at the current estimated age (marked as a solid black line in all panels) to within 1-$\sigma$ of both. The top left panel shows the evolution of the X-ray and EUV luminosities in time by plotting their respective ratios with the bolometric luminosity}. The other five panels show how key characteristics of the planet, and the escape of material from it evolve in time across the three mass loss rate methods we used.
 \label{fig:consist_toi_d}
\end{figure*}

\begin{figure}
\centering
 \includegraphics[width=\columnwidth]{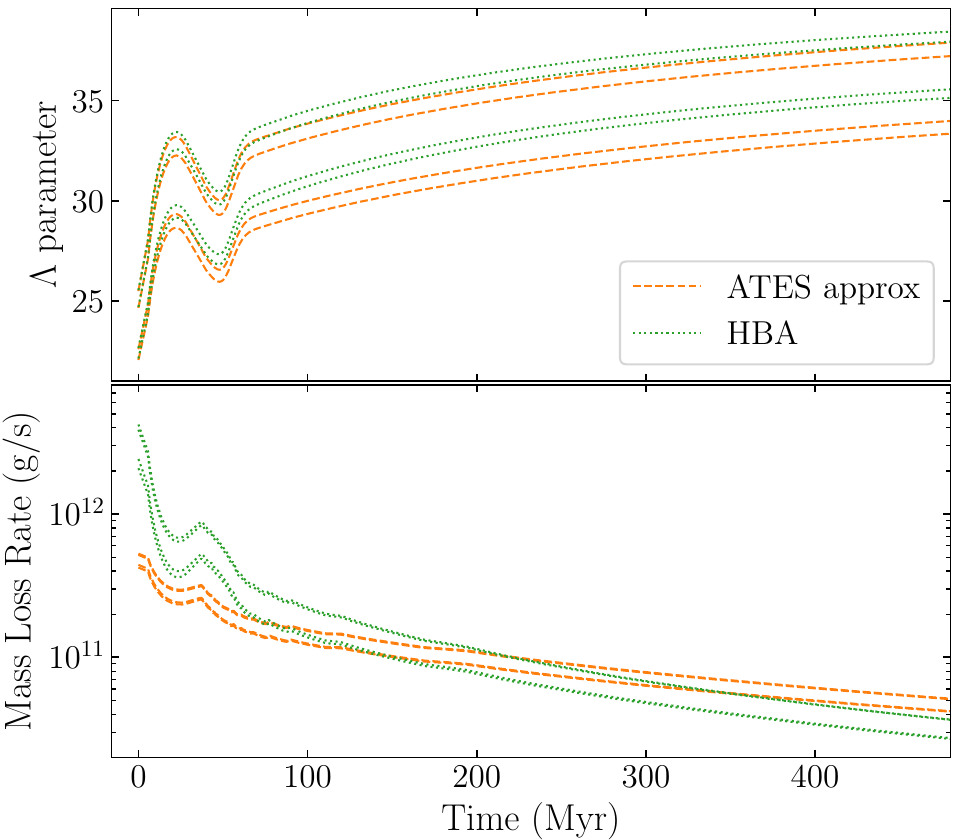}
 \caption{The first 500\,Myr of evolution of the mass loss rate and the restricted Jeans escape parameter, $\Lambda$ (equation 2 of \citealt{Kubyshkina2018HBA}), for the four tracks where both ATES and the hydro-based approximation both reproduce the observed planetary mass and radius within 1-$\sigma$ at the age of the system.}
 \label{fig:consist_early_toi_d}
\end{figure}

Three of the planets in our sample have densities consistent with retaining a substantial gaseous envelope at their current age: TOI-431\,d, \nuLupi\,c, and \nuLupi\,d. We assessed which of the mass loss tracks, described above, resulted in radii and masses that are consistent with the currently measured mass and radius values to within 1-$\sigma$. 

\subsubsection{TOI-431\,d}

In Fig. \ref{fig:heatmap_toi_d} we show a heatmap of the starting grid of masses and envelope mass fractions for TOI-431\,d, highlighting which of those initial conditions led to consistent results for that planet. We label each mass loss rate calculation method, and assign a shade displaying how many of the methods worked. In Fig. \ref{fig:consist_toi_d}, we show the evolution of TOI-431\,d's key characteristics over time for these ``successful" evolutions, as well as the evolution of the high-energy emission from the host star.

The results demonstrate we can successfully reproduce TOI-431\,d with XUV-driven escape, and with all three mass loss methods used. The results do however differ somewhat between the methods. From Fig. \ref{fig:consist_toi_d}, we can see that the energy-limited mass loss rates for the consistent tracks are far lower than the other two methods. This is not to say that there were not other simulations for which the energy-limited method gave higher mass loss rates, or simulations with the other methods which did not give lower rates, but none of them were able to produce TOI-431\,d as we see it today. This is also manifested in the heatmap in Fig. \ref{fig:heatmap_toi_d}, where the successful starting envelope mass fractions for the energy-limited method are below all of the successful values for the other two methods. 

The hydro-based approximation is perhaps at the other extreme. A wide range of starting envelope mass fractions can end up with a TOI-431\,d-like planet at the current age. Looking at Fig.~\ref{fig:consist_toi_d}, it seems that in those cases with much higher starting values, the envelope mass fraction quickly decreases and all of the curves converge to a value less than 10 per cent. 
The calculated mass loss rates that seem to be driving this, as with the energy-limited conclusions above. The early mass loss rates for some tracks are an order of magnitude greater than for ATES, and two orders of magnitude greater than energy-limited, allowing for much greater atmospheric stripping than the other methods. 

To explore this further, in Fig. \ref{fig:consist_early_toi_d} we plot the first 500\,Myr of evolution of the mass loss rate and the restricted Jeans escape parameter, $\Lambda$, from equation 2 of \citet{Kubyshkina2018HBA} for both ATES and hydro-based approximation. For the most direct comparison possible, only the four tracks where both methods reproduced the observed planet mass and radius. This plot demonstrates that the hydro-based approximation mass loss rates are higher than ATES for these planets for about the first 100\,Myr. Some of this could be due to the consideration of the effect of boil-off in the \citet{Kubyshkina2018} models. The initial $\Lambda$ values of 20 lie at the upper limit of where \citet{Kubyshkina2018} expect boil off to be relevant. However, this elevated mass loss rate lasts longer than the expected $\sim$Myr timescale on which boil-off is thought to occur \citep[e.g.][]{OwenWu2016}, and the reason why the rates remains elevated above ATES for much longer is less clear. 
We also note Fig. \ref{fig:consist_toi_d} shows that by a few Gyr in age the mass loss rate for the successful hydro-based approximation tracks are actually all \textit{below} the successful ATES tracks. 

Again, all this is not to say that the other methods did not have simulations with such high mass loss rates early on too, but it seems none of them were able to produce planets with radii and mass consistent with measurements at the current age. 
The fact that the hydro-based approximation is such an outlier suggests that dramatic XUV-driven atmospheric loss, removing up to 25 per cent of the planet's starting mass by the current age of 1.9\,Gyr,
is perhaps in reality unlikely.

\subsubsection{\nuLupi\ c \& d}

\begin{figure*}
\centering
 \includegraphics[width=\textwidth]{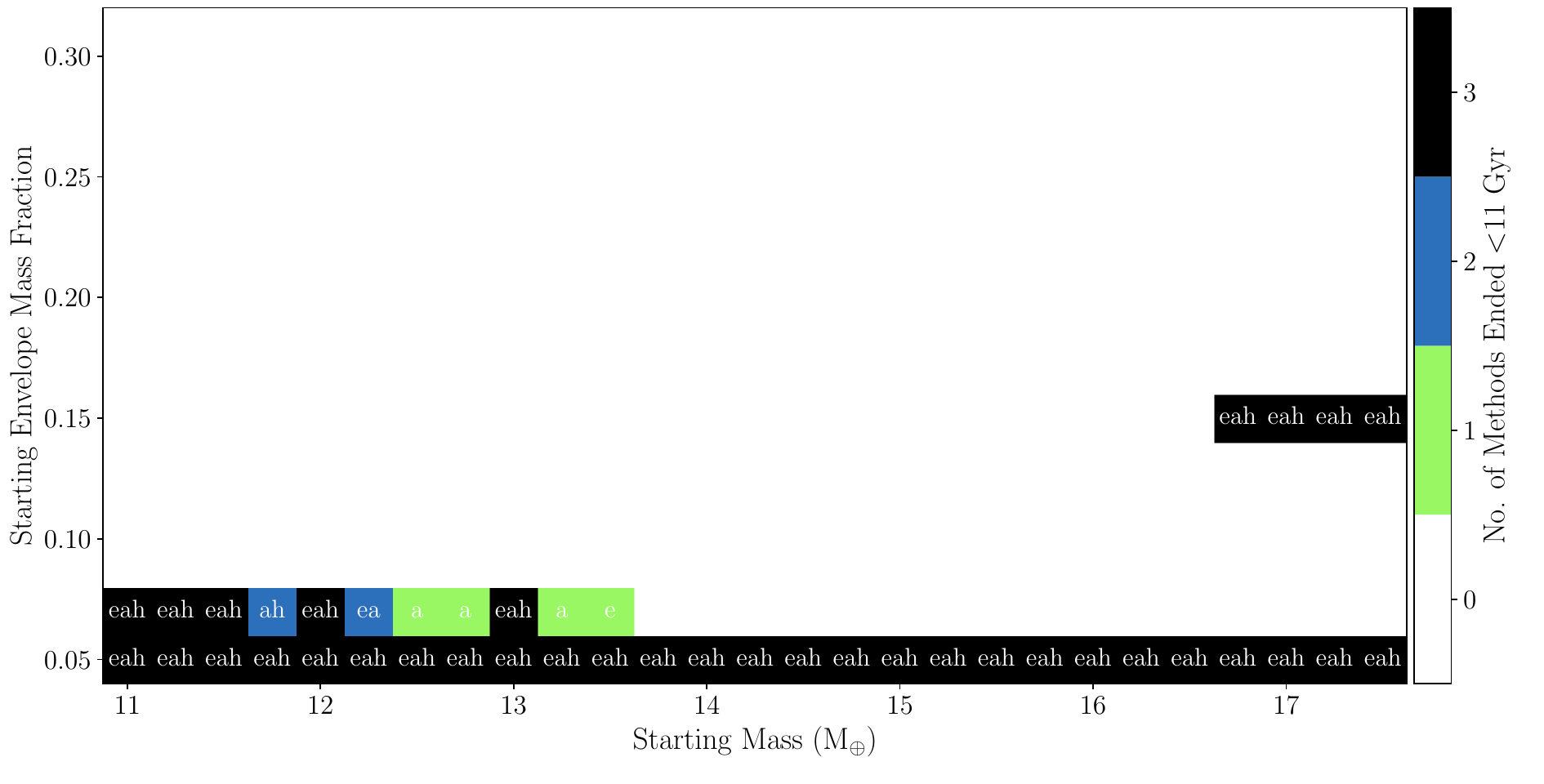}
 \caption{Heat map for the \nuLupi\,c simulations, displaying those evolutionary tracks ended before an age of 11\,Gyr. The squares are coloured according to the number of mass loss calculation methods that suffered an early end to the simulation at that starting point in the grid, while the text labels are as in Fig. \ref{fig:heatmap_toi_d}. We used lower case letters here, as well as a different colour bar, to distinguish from that plot however, as they display very different things.}
 \label{fig:heatmapEarly_nu_c}
\end{figure*}

\begin{figure*}
\centering
 \includegraphics[width=\textwidth]{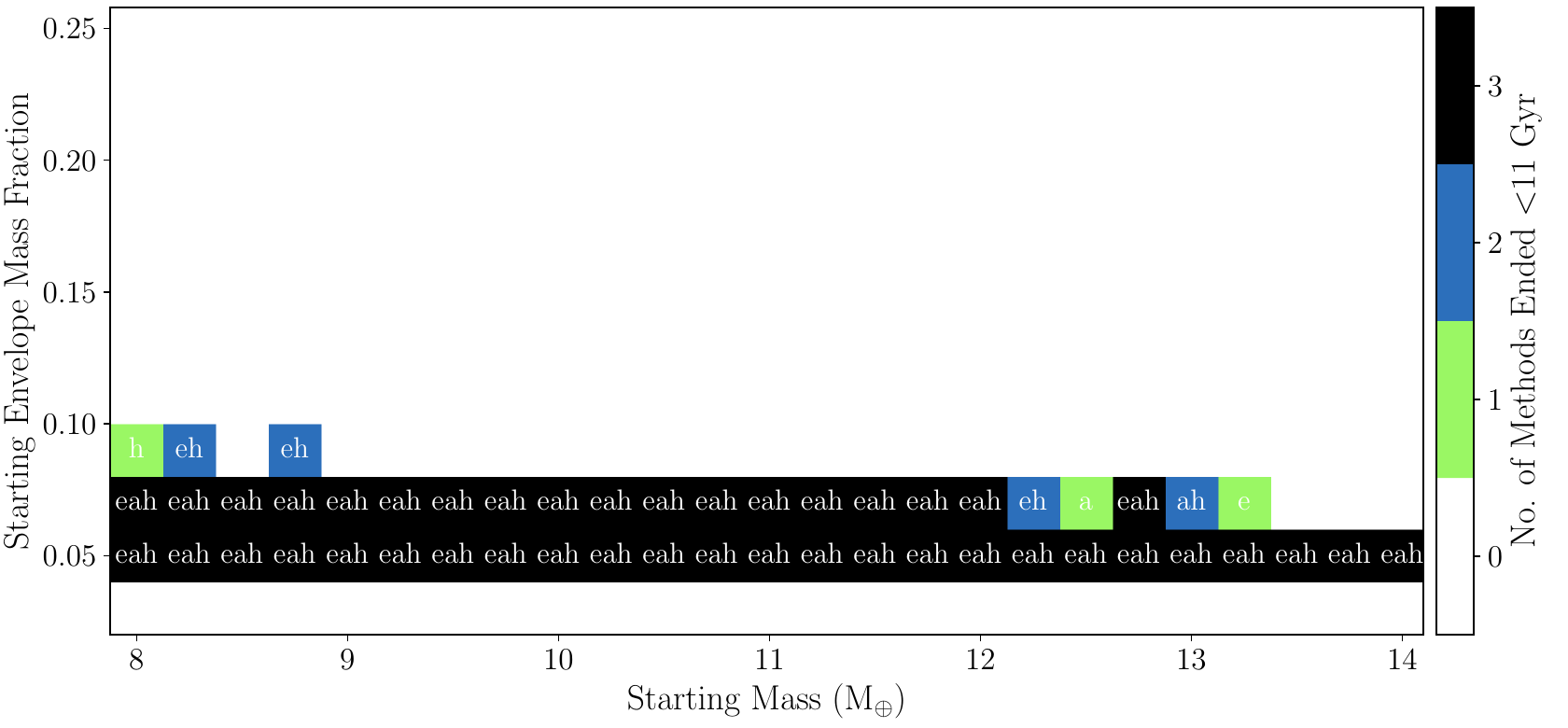}
 \caption{As Fig. \ref{fig:heatmapEarly_nu_c}, but for \nuLupi\,d.}
 \label{fig:heatmapEarly_nu_d}
\end{figure*}

For the other two low density planets in our sample, \nuLupi\,c and d, none of our simulations reproduced the radius at the current age. However, inspection of the results revealed numerous tracks did not reach the system age. For these tracks, MESA was unable to find an acceptable model and so ended the evolution early, often a few Gyr before the 12.3\,Gyr age estimated for the system. Upon investigation, we found that the likely explanation was the evolutionary models reaching the boundaries of what MESA is capable of simulating. The SCVH tables used by MESA in calculating the equation of state at the low densities and temperatures required for these planets hold for $\log Q = \log \rho - 2 \log T + 12 \leq 5.0$ \citep{Saumon1995}. The tracks that ended early were found to have $\log Q_{\rm max} > 5$ in their final few steps before terminating. 
The early end of some simulations can also be seen in some of the consistent results for TOI-431\,d plotted in \ref{fig:consist_toi_d}, where many of the tracks also end before the 10\,Gyr time limit for that planet's simulations. The difference is that TOI-431 is estimated to be much younger, and so we can perform the consistency test at 1.9\,Gyr before the simulations fail later at $\gtrsim 8$\,Gyr.

The key takeaway from the simulations of these two planets is that none of the evolutions 
stripped the planet of its envelope, or was on track to do so. The lack of consistency among those tracks that reached the system age arose because the planet was too large, indicative of an envelope that was too thick, not too thin as to be stripped or nearly so. Moreover, the lowest envelope mass fraction reached in any simulation was 4.69 per cent for planet c, and 4.93 per cent for planet d. Both were for simulations with a starting envelope mass fraction of 0.05, indicating that little mass loss had taken place to reach these values. Given all simulations reached an age well past the early epoch of the highest mass loss rates, it is clear that none of those that ended early would have stripped the planet by the current age, or indeed by the time its star evolves off the main sequence. Even in the absence of our ability to determine consistency at the system age, this is a strong hint that these planets are consistent with the XUV photoevaporation for sculpting the valley.

We can draw a few conclusions about what starting conditions, if any, could have given consistent results for these two planets. In Figs. \ref{fig:heatmapEarly_nu_c} and \ref{fig:heatmapEarly_nu_d}, we show heatmaps of which simulations finished before 11\,Gyr. Since none of those which did successfully run to the system age gave mass and radii that were both consistent, we can rule out any of the grid positions except those highlighted in these heatmaps as successful starting conditions. Almost all of the tracks that ended early have envelope mass fractions below 0.10, in line with that noted by \citet{Malsky2020} when encountering the same issue with the simulation of cooler planets. Additionally, while none of the radii were consistent within 1-$\sigma$ among the simulations that reached the system age, many of the masses were. For both planets, consistent masses were found for starting masses towards the lower end of the grid, with none higher than 12.25\,M$_\oplus$ for planet c, and none higher than 9.75\,M$_\oplus$ for planet d. The dependence of these starting mass consistency limits with mass loss method and starting envelope mass fraction were weak in both cases. This suggests that any consistent results for lower starting envelope mass fraction simulations that did not reach the system age would likely have fallen within limits.

As a further check, we also ran ``hybrid" simulations for these two planets, wherein we evolved the planet for 100\,Myr with MESA, and then the rest of the simulation using the empirical relation from \citet{Chen2016} to recalculate the planet's radius at each timestep. This method yielded consistent masses and radii at the system age for a few of the simulations for each planet: for c, four simulations all starting at an envelope mass fraction of 0.07, with masses starting between 11.0 and 11.75\,M$_\oplus$; for d, eight simulations all starting at an envelope mass fracton of 0.05, and starting masses between 8.0 and 9.75\,M$_\oplus$. These starting conditions are in line with what we inferred above from MESA alone.

\subsection{Results for Stripped Planets}
\label{ssec:stripped}

 





\begin{figure}
\centering
 \includegraphics[width=0.8\columnwidth]{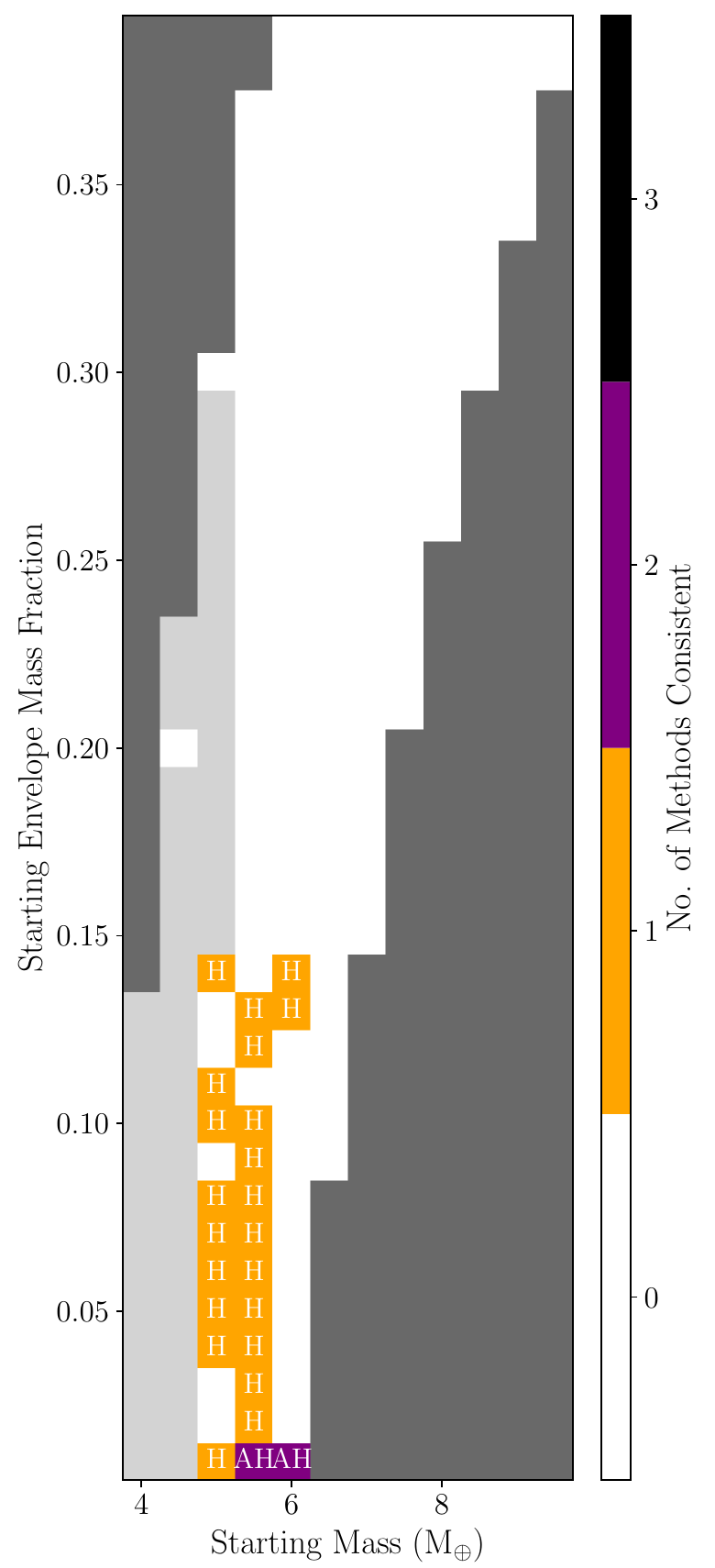}
 \caption{Heat map for the \nuLupi\,b, displaying the starting conditions which result in evolutionary tracks where the radius and mass are consistent within the measured values to within 3-$\sigma$ at the closest timestep to the current estimate age. Note that in some cases, the tracks end many Gyr before the estimated age. Light grey regions are those where MESA failed to reach the evolution step. Dark grey regions in the top left and bottom right are excluded as they imply a core mass that is not within 3-$\sigma$ of the current measured mass, as we assume the core mass is unchanged during the evolution. The text labels indicating the consistent models are as in Fig. \ref{fig:heatmap_toi_d}.}
 \label{fig:heatmap_nu_b_consist}
\end{figure}

\begin{figure*}
\centering
 \includegraphics[width=\textwidth]{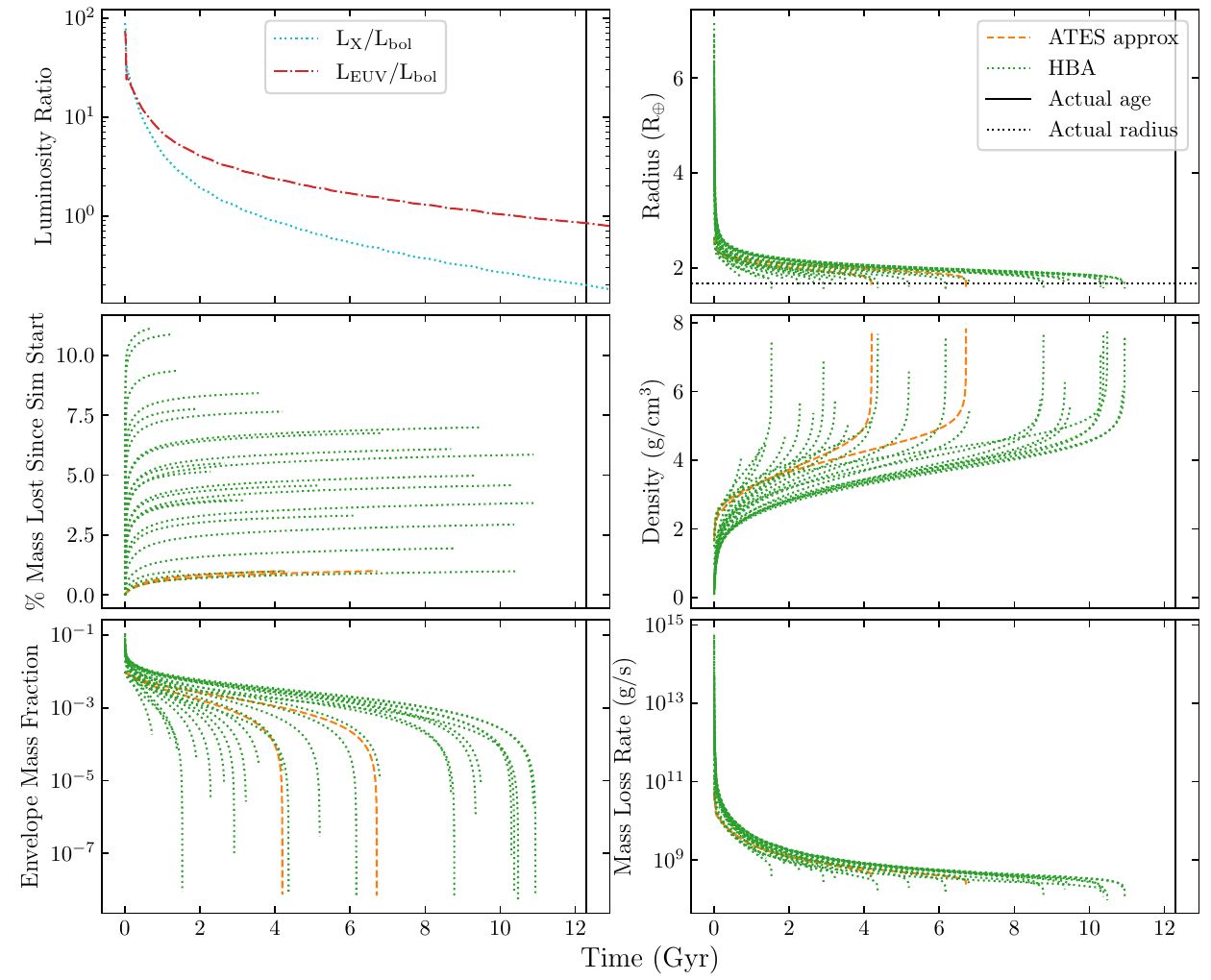}
 \caption{The subset of evolutionary tracks for \nuLupi\,b where the planet's radius (marked as a black dotted line in the top right panel) and mass matched the measured values at the current estimated age (marked as a solid black line in all panels) to within 3-$\sigma$ of both. The layout and labels are as in Fig. \ref{fig:consist_toi_d}.}
 \label{fig:consist_nu_b}
\end{figure*}

\begin{figure}
\centering
 \includegraphics[width=0.8\columnwidth]{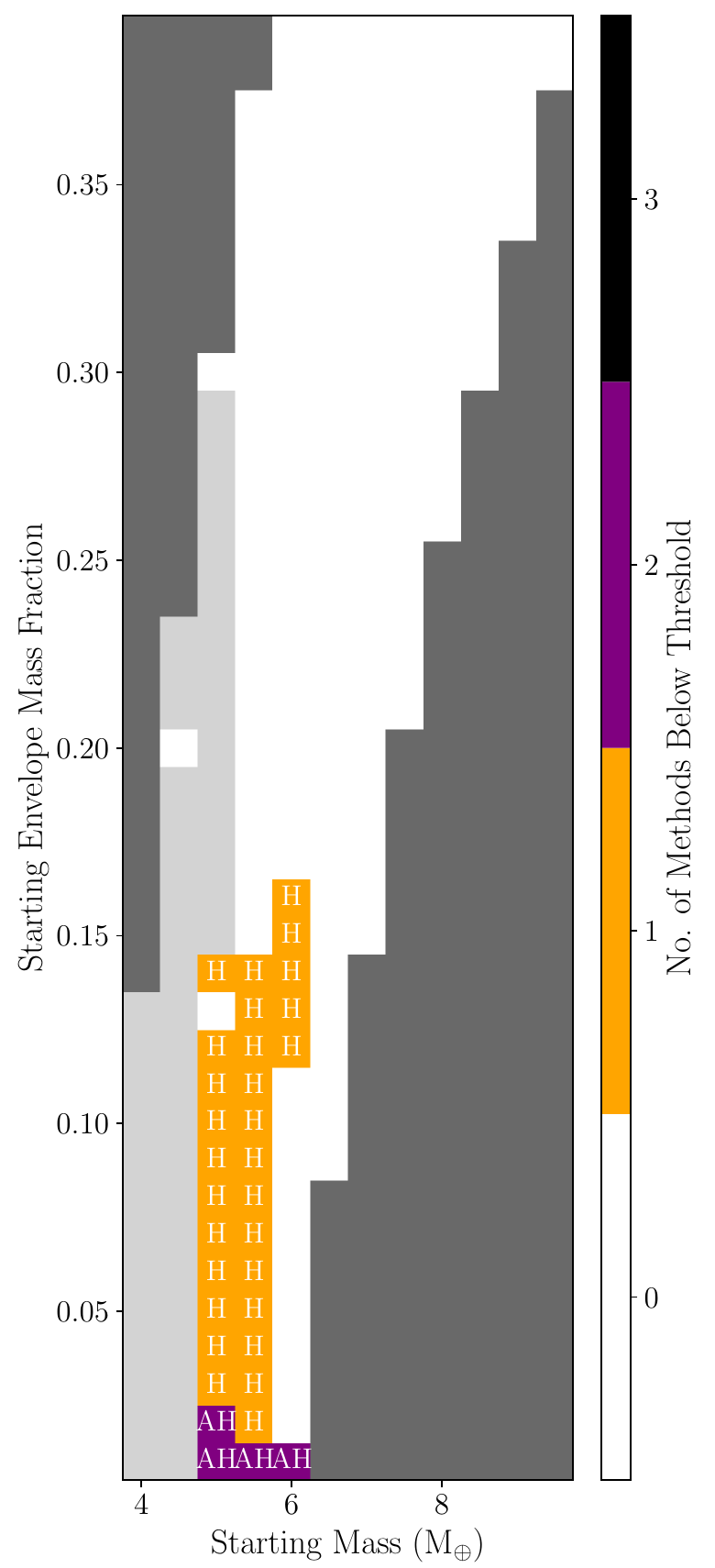}
 \caption{As Fig. \ref{fig:heatmap_nu_b_consist} but showing all tracks whose final envelope mass fraction was below 0.0002 (see main text), regardless of the age of the stopping point of the simulation.}
 \label{fig:heatmap_nu_b_threshold}
\end{figure}

TOI-431\,b and \nuLupi\,b both have small radii and high densities, and are therefore consistent with already having been stripped of any primordial envelope they may have had. In this instance, we ran the simulations over a wider grid. 
We explored the limits of what initial conditions would lead the planet to lose its atmosphere. 
In these simulations, we assumed that any atmosphere of heavier elements that remains now increases the mass and radius by a negligible amount. Therefore, we only allowed starting conditions that implied an initial core mass that is within 3-$\sigma$ of the current measured planetary mass. 

\subsubsection{TOI-431\,b}

None of the simulations for TOI-431\,b advanced more than a few Myr in the maximum 5000 steps we initially allowed for the evolution. The timesteps in MESA are dynamic, and so we interpreted the code forcing the them to be small being due to rapid changes to the planet over short periods of time. We further investigated by running a subsample of the original grid for 200,000 steps, which also failed to advance beyond the first few Myr after our starting age. Inspection of the tracks revealed very large initial planetary radii ($>50$\,R$\oplus$) and brisk further expansion across the small elapsed time the simulations were able to cover.

The failure of our simulations for this planet are likely reflective of the extreme environment in which this planet resides, with an orbital separation of just 0.011\,au, and its relatively low total mass. In the final setup stage for our models in MESA before the planet is evolved with XUV-driven escape, the planet is gradually irradiated to the level of the bolometric flux of the star, calculated from its $T_{\rm eff}$, over 6\,Myr \citep{Malsky2020} The small orbital separation therefore leads to this final setup stage puffing up the planet significantly under the intense bolometric irradiation, and when we attempt to then evolve the planet under XUV irradiation, it undergoes further rapid expansion.

In lieu of usable results from our simulations, we can still make some inferences if we assume some reasonable characteristics for a TOI-431\,b planet. We took a starting mass of 4\,M$_\oplus$, close to the centre of our test grid. We then also assumed a bulk density of 0.5\,g\,cm$^{-3}$ (implying an initial radius of 3.5\,R$\oplus$); a conservatively high initial value when compared to the simulations of \nuLupi\,b in Section \ref{sssec:nuLupib}, where the initial densities simulated by MESA are much lower. We calculated the XUV flux at the planet from the first Mors step for TOI-431 ($4.4\times10^6$\,erg\,s$^{-1}$\,cm$^{-2}$), leading to an energy-limited escape mass loss rate of $2\times10^{14}$\,g. This is sufficient to remove a 20 per cent initial envelope in 0.6\,Myr, assuming a constant mass loss rate. Therefore, while we cannot simulate the system in MESA, this quick calculation demonstrates the ease with which a primordial envelope may have been removed, due the short orbital separation.

\subsubsection{\nuLupi\,b}
\label{sssec:nuLupib}

The results for \nuLupi\,b are very different to TOI-431\,b. In the majority of the simulations, the MESA evolution made it to the estimated system age. We tested the simulated tracks for consistency with the measured mass and radius, but loosened the restrictions as compared to the tests for gaseous planets. Firstly, we tested the closest point in the track to the system age, regardless of if the track ended early. We did not want to skip planets stripped well before 12.3\,Gyr, not least because their radius and mass have likely changed little between the point the envelope was removed entirely and the present day (unlike gaseous planets for which the track ended early with an appreciable atmosphere remaining). Second, we only required that the mass and radius be within 3-$\sigma$ of the measured values, rather than 1. Our primary aim with these simulations is investigating whether the planet can be stripped by XUV-driven escape, as opposed to reproducing the remnant planet core exactly.

In Figs. \ref{fig:heatmap_nu_b_consist}, we display a heatmap of the starting conditions which resulted in planet with mass and radius consistent with \nuLupi\,b within 3-$\sigma$, similar to \ref{fig:heatmap_toi_d} for TOI-431\,d. There are two extra grey regions on this plot. The dark grey shows combinations of the starting conditions that implied a core mass more than 3-$\sigma$ different from the measured mass today, and were thus excluded from our analysis. The light grey region shows where MESA failed to reach a stable solution in the earlier steps of setting the planet model up (which uses existing code that we did not edit in our study), and so the evolution under XUV irradiation was not able to be tested. 
For the tracks that were found to be consistent, we also plotted their evolution of the key planetary characteristics in Fig. \ref{fig:consist_nu_b}.

The vast majority of the consistent tracks were found using the hydro-based approximation, while a few were found with ATES. Starting masses between 5 and 6\,M${\oplus}$ were favoured, although almost all between 4 and 5\,M${\oplus}$ failed to find a stable solution in setting up the model in MESA and so we not able to be tested. The successful starting envelope mass fractions were over a wider range, with viable simulations for all values tested below 0.15, although both successful ATES tracks started at 0.01, the lowest initial value. There is no obvious relationship between starting envelope mass fraction and starting mass in terms of which combinations give consistent results. Energy-limited escape was not able to reproduce the planet for any starting conditions we evolved under XUV-driven escape. However, with the starting masses between 4 and 5\,M${\oplus}$ largely unable to be tested, we cannot completely rule out the possibility of the method being able to reproduce \nuLupi\,b. This is again though suggestive of the method underestimating mass loss rates as compared to those based on the hydrodynamic models, such as the other two methods we employ here.

As an additional test of which starting conditions could lead to a stripped planet, we also investigated simply setting a threshold envelope mass fraction at the end of the simulations. To determine this threshold, we examined the last step of each of the consistent tracks within 3-$\sigma$ from above, and found the highest envelope mass fraction among them: 0.0002. If we assume anything below this is stripped (or imminently will be), a few extra starting conditions result in a stripped planet, as shown in Fig. \ref{fig:heatmap_nu_b_threshold}. The typical ranges of the two starting conditions is largely unchanged, with the extra ``successful" tracks in this case most filling in gaps within the consistent region from Fig. \ref{fig:heatmap_nu_b_consist}. The energy-limited method still fails to remove the envelope of the planet models under this threshold definition of stripping.

Finally, we tested what happened if we adopted an XUV history of \nuLupi\ that was at the median level for a star of its mass. The median Mors track suggests the star was under-bright in X-rays for its age in our observations. With four test cases that did not strip for any method in our main results (pairings of starting masses of 6, 6.5, 7.0, and 7.5 with starting envelope mass fractions of 0.05, 0.10, 0.15, and 0.21, respectively). The hydro-based approximation strips all four of these cases, while ATES strips all but the highest mass case. Energy-limited still fails to reproduce the stripped core composition. In general, these results suggest that if \nuLupi\ happened to be at a lower than average brightness during our observations, then many more starting conditions may have been possible than suggested by Fig \ref{fig:heatmap_nu_b_consist}.

\subsection{Current Mass Loss rates}
\label{ssec:currentML}

\begin{table*}
\centering
\caption{Estimates of the current mass loss rate for the three transiting planets in our sample consistent with being gaseous, using four methods: energy-limited escape ($\dot{M}_{\rm En}$), the full \textsc{Ates} code ($\dot{M}_{\rm ATES}$, \citealt{Caldiroli2021}), the approximate effective efficiency function \citep[$\dot{M}_{\rm AEEF}$,][]{Caldiroli2022}, use of the interpolation tool across the \citet{Kubyshkina2018} grid of models ($\dot{M}_{\rm Int}$) \citep[see also][]{Kubyshkina2021}, and the hydro-based approximation \citep[$\dot{M}_{\rm HBA}$][]{Kubyshkina2018HBA}. We also give the measured radii and masses for each planet, taken from \citet{Osborn2021} and \citet{Delrez2021} for TOI-431 and \nuLupi, respectively.}
\label{tab:currentML}
\begin{threeparttable}
\begin{tabular}{lccccccc}
\hline
Planet     & $R_{\rm p}$                & $M_{\rm p}$            & $\dot{M}_{\rm En}$ & $\dot{M}_{\rm ATES}$ & $\dot{M}_{\rm AEEF}$& $\dot{M}_{\rm Int}$& $\dot{M}_{\rm HBA}$  \\
         & R$_{\oplus}$               & M$_{\oplus}$             & g\,\ps             & g\,\ps             & g\,\ps              & g\,\ps     & g\,\ps               \\ \hline
         \\[-0.25cm]
TOI-431\,d & $3.29^{+0.09}_{-0.08}$    & $9.90^{+1.53}_{-1.49}$  & $1.5\times10^{9}$  & $7.8\times10^{9}$  &  $8.8\times10^{9}$  & $6.3\times10^{9}$  & $3.6\times10^{9}$  \\[0.05cm]
\nuLupi\,c & $2.916^{+0.075}_{-0.073}$ & $11.24^{+0.65}_{-0.63}$ & $6.1\times10^{7}$  & $4.6\times10^{8}$  & $3.8\times10^{8}$  & $7.3\times10^{8}$  & $1.4\times10^{8}$  \\[0.05cm]
\nuLupi\,d & $2.562^{+0.088}_{-0.079}$ & $8.82^{+0.93}_{-0.92}$  & $8.4\times10^{6}$  & $\dagger$          & $4.6\times10^{7}$  & $3.5\times10^{7}$  & $8.3\times10^{6}$  \\[0.05cm] \hline
\end{tabular}
\begin{tablenotes}
\item $\dagger$ The \textsc{Ates} code did not successfully converge for \nuLupi\,d (see main text).
\end{tablenotes}
\end{threeparttable}
\end{table*}

In Table \ref{tab:currentML} we give current mass loss rates of the three transiting planets in our sample that have densities consistent with them being gaseous. We compared the different mass loss rate estimation methods used in Section \ref{sec:sims}, along with two extra methods: the interpolator for the \citet{Kubyshkina2018} models \citep{Kubyshkina2021}, and running the full hydrodynamic \textsc{Ates} code \citep{Caldiroli2021}. For \nuLupi\,d, the code failed to converge in a reasonable amount of time, and so we do not give a value for this method. The failure to converge may be linked to its current low irradiation level of $14.1^{+3.0}_{-3.4}$\,\flux, meaning the planet may be in the Jeans escape regime, and thus modelling it with hydrodynamic escape at current age may be inappropriate.

We did not calculate current mass loss rates for TOI-431\,b and \nuLupi\,b because our mass loss rate methods all assume a hydrogen-dominated atmosphere. These two planets are thought to not retain any envelope and thus the applicability of these methods at the current epoch is questionable.

Looking at Table \ref{tab:currentML}, we can compare the results across the different methods, particularly those based on the same models. The mass loss rates given by the hydro-based approximation, $\dot{M}_{\rm HBA}$, are all systematically smaller than those given by the interpolator tool, $\dot{M}_{\rm Int}$, by factor of between two and five. From fig. 3 of \citet{Kubyshkina2018HBA}, this factor appears to be in line with expectation. Overall, one would expect the interpolator to be a better approximation to the \citet{Kubyshkina2018} models, as it provides a local interpolation to that region of the grid, as opposed to a single analytic expression across all models. Meanwhile, those calculated with the \textsc{Ates} code, $\dot{M}_{\rm ATES}$, are encouragingly matched very well by the analytic fit to those models: the approximate effective efficiency function, $\dot{M}_{\rm AEEF}$. Finally, all five methods agree that TOI-431\,d is undergoing the most mass loss currently, and \nuLupi\,d the least.

\section{Testing the systems with EvapMass}
\label{ssec:EvapMass}

As both TOI-431 and \nuLupi\ both have planets above and below the radius valley, we used the \textsc{EvapMass} code \citep{Owen2020} as an additional test of the systems' compatibility with the photoevaporation mechanism. This code takes the parameters of a rocky planet in the system and calculates the minimum mass of a second, gaseous planet to be consistent with the photoevaporation mechanism. The code exploits the fact that the two planets have experienced the same XUV emission from the star, scaled to their respective orbital distances. In their study, \citet{Owen2020} explored 73 systems to determine if they are consistent with the photoevaporation model, finding only two systems that appear inconsistent.

For TOI-431, we performed the test with planets b and d as the stripped and gaseous planets, respectively, and found them to be consistent with photoevaporation. However, the constraint provided is not very informative. Planet b is so small and close to the star that it provides very little constraint on planet d. The minimum mass required for consistency is just 0.099\,M$_\oplus$ at 95 per cent confidence, compared to the true mass of planet d which is $9.9\pm1.5$\,M$_\oplus$.


With \nuLupi, we applied the code to two planet pairs: b and c, and b and d. 
Planet d's required minimum mass is a similarly low, 0.59\,M$_\oplus$ at 95 per cent confidence. For planet c, the constraint is somewhat more informative, with a minimum mass of 2.7\,M$_\oplus$. However, the measured mass of this planet is again far above this minimum confidence limit, at $11.24^{+0.65}_{-0.63}$\,M$_\oplus$.

This method demonstrates that the two systems in this work are consistent with the photoevaporation model, in agreement with the results of our simulations in Section \ref{sec:sims}, though its minimum mass constraints are not strong. The \textsc{EvapMass} code is more informative for systems with a gaseous planet interior to the stripped one. One such example is K2-3, for which the measured parameters are not consistent with photoevaporation, according to the minimum mass limits calculated by the code \citep{HDL2022}.

\section{Conclusions}

We have examined the TOI-431 and \nuLupi\ planetary systems, both of which host planets either side of the radius valley, in the context of XUV photoevaporation. Using \textit{XMM-Newton} EPIC observations of the systems, we measured the stars' X-ray fluxes. These measurements will prove useful in interpreting any future observations targeting the detection of escaping or extended atmospheres of the planets in these systems, such as at Ly-$\alpha$. The X-ray and NUV light curves for TOI-431 show evidence of a possible stellar flare with a rise in both at the end of the observation. There is, however, no evidence of the emission hardening at that epoch, meaning that we cannot conclusively determine if the increase in X-ray and UV flux can definitively be attributed to a stellar flare.

Using the current X-ray flux as an anchor for the XUV evolution of each host star, we simulated the lifetime evolution of a H/He envelope for each of the five transiting planets across the two systems. Building on the work of \citet{PraesepePaper}, we simulated the past evolution as well as the future, starting at an age of 5\,Myr. For the XUV time evolution, we generated evolution curves with the \textsc{Mors} code \citep{Johnstone2021}, and anchored them to the XUV fluxes we measured with the \textit{XMM-Newton} observations. To get the planet structure, and therefore radius, at each step, we used MESA to evolve our planets, building on previous work by \citep[e.g.][]{Chen2016,Malsky2020}. We used multiple methods of estimating the mass loss rate throughout the observations, comparing the predicted mass and radius to those observed today.

Our results demonstrate TOI-431\,d can be easily reproduced by all mass loss methods we employed, though the viable starting conditions varied between these methods. Energy-limited favoured a smaller starting envelope mass fraction of 0.07, while the hydro-based approximation to the \citet{Kubyshkina2018} models suggested it could be as high as 0.25, though most successful models were found just either side of 0.1. Our simulations for \nuLupi\,c and d encountered issues with the equation of state tables used by MESA, causing numerous tracks to terminate prior to the older system age of 12.3\,Gyr. However, none of the simulations were able to strip the envelope off either planet, or were on track to do so. We conclude that the survival of their envelopes to the present day, as shown by their measured densities, is consistent with the XUV-driven escape model.


Our simulations for TOI-431\,b also encountered issues due to high bolometric flux at the planet in the final setup stage, leading to very large radii. However, a quick calculation suggests that removing a 20 per cent initial primordial envelope could be achieved on Myr timescales due to the very small orbital separation and thus intense XUV irradiation the planet would have experienced immediately following disk dissipation.

Our simulations were also able to completely remove the primordial envelopes of \nuLupi\,b, for at least some combinations of starting mass and envelope mass fraction. Complete stripping is in line with expectations from its measured density. 
While none of the energy-limited method simulations resulted in a complete removal, the other two methods were able to strip the planet in some cases, Furthermore, there were a number of lower starting mass simulations that failed in MESA before we reached the XUV evolution stage, and so were not able to be tested.

Our simulations, along with use of the \textsc{EvapMass} code \citep{Owen2020}, show that XUV-driven escape can successfully reproduce the system architectures seen, with the innermost planet stripped, and the outer transiting planets able to retain at least some of their primordial envelope. These systems therefore appear to be consistent with the XUV mechanism for forming the radius valley. Whether it is the dominant mechanism, for these systems or indeed in general, remains an open question.

\section*{Acknowledgements}
We thank Andrea Caldiroli and Riccardo Spinelli for their help with the \textsc{Ates} code, and James Owen for helpful discussions. This work is based on observations obtained with {\it XMM-Newton}, an ESA science mission with instruments and contributions directly funded by ESA Member States and NASA. These observations are associated with \textit{XMM-Newton} OBSIDs 0884680101 and 0884680201. This work was supported by NASA under award No 80NSSC22K0795.

\section*{Data Availability}
The \textit{XMM-Newton} data used in this work are publicly available from the \textit{XMM-Newton} Science Archive: \url{http://nxsa.esac.esa.int/nxsa-web/#search}.



\bibliographystyle{mnras}
\bibliography{radiusvalley} 

\bsp	
\label{lastpage}
\end{document}